\documentclass[letterpaper]{article} 
\usepackage{aaai25}  
\usepackage{times}  
\usepackage{helvet}  
\usepackage{courier}  
\usepackage[hyphens]{url}  
\usepackage{graphicx} 
\usepackage{braket}
\usepackage{tikz}
\usepackage{quantikz}
\usepackage{amsmath}
\usepackage{amsfonts} 
\usepackage{tensor} 
\usepackage{float}
\usepackage{tabularx}

\usepackage{enumitem}

\usepackage{natbib}  
\usepackage{caption} 
\usepackage{algorithm}
\usepackage{algorithmic}

\usepackage[utf8]{inputenc} 
\usepackage[T1]{fontenc}    
\usepackage{url}            
\usepackage{booktabs}       
\usepackage{amsfonts}       
\usepackage{nicefrac}       
\usepackage{microtype}      
\usepackage{xcolor}         
\usepackage{graphicx}
\usepackage{amsmath}
\usepackage{newfloat}
\usepackage{listings}
\DeclareCaptionStyle{ruled}{labelfont=normalfont,labelsep=colon,strut=off} 
\floatstyle{ruled}
\newfloat{listing}{tb}{lst}{}
\floatname{listing}{Listing}
\title{Reinforcement Learning for Quantum Circuit Design:\\ Using Matrix Representations}

\author {
    Zhiyuan Wang,
    Chunlin Feng,
    Christopher Poon,
    Lijian Huang, \\
    Xingjian Zhao,
    Yao Ma, Tianfan Fu, Xiao-Yang Liu
}
\affiliations {
    Department of Computer Science, Rensselaer Polytechnic Institute \\

    Emails: \{wangz60, fengc5, poonc3, huangl9, zhaox8,  may13, fut2, liux33\}@rpi.edu
}

\begin{document}

\maketitle

\begin{abstract}

Quantum computing promises advantages over classical computing. The manufacturing of quantum hardware is in the infancy stage, called the Noisy Intermediate-Scale Quantum (NISQ) era. A major challenge is automated quantum circuit design that map a quantum circuit to gates in a universal gate set. In this paper, we present a generic MDP modeling and employ Q-learning and DQN algorithms for quantum circuit design. By leveraging the power of deep reinforcement learning, we aim to provide an automatic and scalable approach over traditional hand-crafted heuristic methods.
\end{abstract}

\section{Introduction}

Quantum computing has the potential to revolutionize computing beyond the reach of classical computers \cite{gill2021quantumcomputingtaxonomysystematic}. A major hurdle is the quantum circuit design that maps a quantum circuit to gates in a universal gate set. Traditional hand-crafted heuristic methods are often inefficient and not scalable. 

The automated design of quantum circuits remains a major challenge. \cite{ali2015quantum} and \cite{bhat2022optimal} used a method that utilizes the Toffoli gate decomposition technique, reducing cost and enhancing efficiency. 
Machine learning, in particular reinforcement learning, has recently been applied. \cite{sogabe2022model} explored a model-free deep recurrent Q-network (DRQN) method for an entangled Bell-GHZ circuit. Recently,  \cite{NEURIPS2023_d41b7001} and \cite{meirom2022optimizing} utilized tensor network representations of Google's Sycamore circuit \cite{arute2019quantum} and studied the Tensor Network Contraction Ordering (TNCO) problem.

In this paper, we explore reinforcement learning methods to automate the task of quantum circuit search. Our contributions can be summarized as follows:
\begin{itemize}[leftmargin=*]
    \item We present three generic Markov Decision Process (MDP) modelings for the quantum circuit design task.
    \item We study $10$ quantum circuit design tasks: $4$ Bell states, SWAP gate, iSWAP gate, CZ gate, GHZ gate, Z gate and Toffoli gate, respectively, given a universal gate set \{$H$, $T$, $\text{CNOT}$\}. 
    \item We verify that both Q-learning and DQN algorithms could find the target quantum circuits. Reinforcement learning offers an automated solution. 
\end{itemize}




\section{Problem Formulation}

Taking Bell state $\left|\Phi^+\right\rangle$ \footnote{There are four Bell states, physically the two qubits are maximumly entangled.} as an example, we formulate the task of quantum circuit design as two versions of Markov Decision Process (MDP). In particular, we specify the state space, action set, reward function, and Q-table, respectively.

\subsection{Task: Quantum Circuit Design}

\begin{figure}[t]
\centering
\begin{tikzpicture}
    \node (circuit) [inner sep=0pt] {
        \begin{quantikz}
            q_0:~ \ket{0} & \gate{H} & \ctrl{1} & \qw \\
            q_1:~ \ket{0} & \qw & \targ{} & \qw
        \end{quantikz}
        
    };
    \node at ([yshift=-1cm,xshift = 1.2cm]circuit) {CNOT$_{01}$};
    \node[right=0cm of circuit,yshift=-0.1cm] {$\ket{\Phi^+} = \frac{1}{\sqrt{2}}(\ket{00} + \ket{11})$};
\end{tikzpicture}
\caption{A quantum circuit to generate Bell state $\ket{\Phi^+}$.}
\label{fig:bell_state_0}
\vspace{-0.2in}
\end{figure}
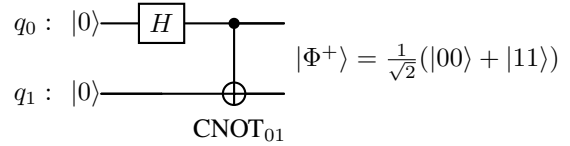

Given two qubits with initial state $|q_1 q_0\rangle = |00\rangle$ and a universal gate set $G = \{H, T, \text{CNOT}\}$, the goal is to find a quantum circuit that generates the Bell state $\left|\Phi^+\right\rangle$: 
\begin{equation}\label{eq:target_state111}
    \left|\Phi^+\right\rangle = \frac{1}{\sqrt{2}} \left( \left|00\right\rangle + \left|11\right\rangle \right).
\end{equation}

The target quantum circuit to generate $\left|\Phi^+\right\rangle$ is shown in Fig. \ref{fig:bell_state_0}, whose matrix representation is:
\begin{equation}\label{eq:target_matrix}
\begin{split}
U &= \text{CNOT}_{01} \cdot (H \otimes I) \\
&=
  \begin{pmatrix}
  1 & 0 & 0 & 0 \\
  0 & 1 & 0 & 0 \\
  0 & 0 & 0 & 1 \\
  0 & 0 & 1 & 0 
  \end{pmatrix}
  \cdot \left(
  \frac{1}{\sqrt{2}}
  \begin{pmatrix}
      1 & 1\\
      1 & -1
  \end{pmatrix}
  \otimes
  \begin{pmatrix}
      1 & 0\\
      0 & 1\\
  \end{pmatrix} \right)\\
  &=
  \frac{1}{\sqrt{2}}
  \begin{pmatrix}
  1 & 0 & 1 & 0 \\
  0 & 1 & 0 & 1 \\
  0 & 1 & 0 & -1 \\
  1 & 0 & -1 & 0 
  \end{pmatrix}.
\end{split}
\end{equation}
Note that $\left|\Phi^+\right\rangle = U~ |00\rangle$.

\subsection{Modeling as Markov Decision Process (MDP)}

We provide three types of MDP modelings. 

\subsubsection{Matrix Representation}

\begin{figure}
    \centering
    \includegraphics[width=0.5\textwidth]{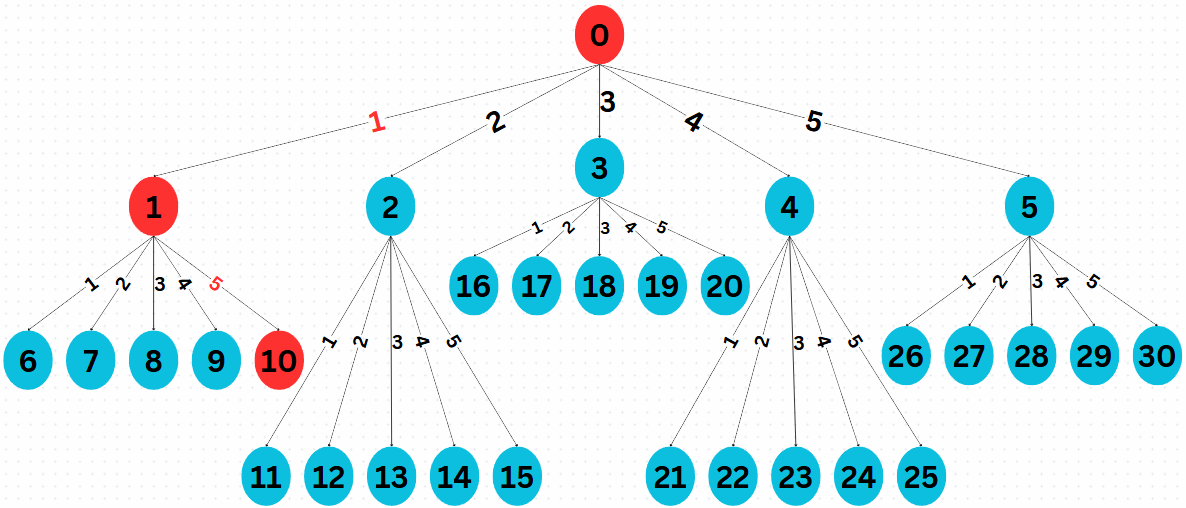}
    \caption{State tree in matrix representation for searching the circuit in Fig. \ref{fig:bell_state_0}.}
    \label{fig:bell_state_1}
    \vspace{-0.2in}
\end{figure}

\begin{itemize}[leftmargin=*]
\item Actions
$\mathcal{A} = \{H_0,H_1,T_0,T_1,\text{CNOT}_{01}\}$, since $H$ and $T$ can be executed on either $q_0$ or $q_1$. An action $a \in \mathcal{A}$ is represented as a matrix \(\mathit{A}\in \mathbb{C}^{4\times 4}\). 
\item State space $\mathcal{S}$: The initial state is $U_0 = I_{4}$ and the terminal state is $U$ given in (\ref{eq:target_matrix}). 
Let $S$ be the current state (a node in Fig. \ref{fig:bell_state_1}), $A \in \mathcal{A}$ be the action, then the resulting state at a child node $S^{'}$ is given by
\begin{equation}
   S^{'} = A \cdot S. 
\end{equation}
The state space $\mathcal{S}$ is a tree in Fig. \ref{fig:bell_state_1}. The connecting lines 1, 2, 3, 4, and 5 correspond to the five actions in $\mathcal{A}$. At the initial state $S_0 = I_4$, taking an action $a \in \mathcal{A}$ will generate $5$ states $\{S_1, S_2, S_3, S_4, S_5\}$. Then, taking a second action $a \in \mathcal{A}$ at a state $S \in \{S_1, S_2, S_3, S_4, S_5\}$ will generate $25$ states $\{S_6, S_7, \ldots, S_{30}\}$. Thus, $\mathcal{S}$ has a total of $31$ states.

\item Reward function $R$: At state $S_1
$, taking action $a = \text{CNOT}_{01}$, the reward is $R(s = S_1, a = \text{CNOT}_{01})= 100$; otherwise, $R(s, a)=0$. 
\end{itemize}



\noindent\textbf{Example for Fig. \ref{fig:bell_state_0}}: Given initial state $S_0 = I_4$, let us consider the optimal trajectory $S_0 \rightarrow S_1 \rightarrow S_{10}$.

  
\noindent\textbf{State after taking the first action} $a = H_0$,
\begin{equation}
\begin{split}
S_1 = (H_0 \otimes I) S_0
   =\frac{1}{\sqrt{2}}
  \begin{pmatrix}
  1 & 0 & 1 & 0 \\
  0 & 1 & 0 & 1 \\
  1 & 0 & -1 & 0 \\
  0 & 1 & 0 & -1 
  \end{pmatrix}.
\end{split}
\end{equation}
  
\noindent\textbf{Final state after taking the second action} $a = \text{CNOT}_{01}$,
\begin{equation}
\begin{split}
S_{10} &= \text{CNOT}_{01} \cdot S_1 \\
& =\frac{1}{\sqrt{2}} \begin{pmatrix}
  1 & 0 & 0 & 0 \\
  0 & 1 & 0 & 0 \\
  0 & 0 & 0 & 1 \\
  0 & 0 & 1 & 0 
  \end{pmatrix}
  \begin{pmatrix}
  1 & 0 & 1 & 0 \\
  0 & 1 & 0 & 1 \\
  1 & 0 & -1 & 0 \\
  0 & 1 & 0 & -1 
  \end{pmatrix} = U,
\end{split}
\end{equation}
which corresponds to the target circuit in (\ref{eq:target_matrix}).


  \textbf{Advantage}: Different sequences of quantum gates may result in the same matrix state, thus this matrix representation would reduce the state space.
  
  \textbf{Disadvantage}: RL agent needs to be trained for each target matrix, even though different circuits may share similar or identical intermediate states. This approach makes the training process repetitive.


\begin{figure}
    \centering
    \includegraphics[width=0.5\textwidth]{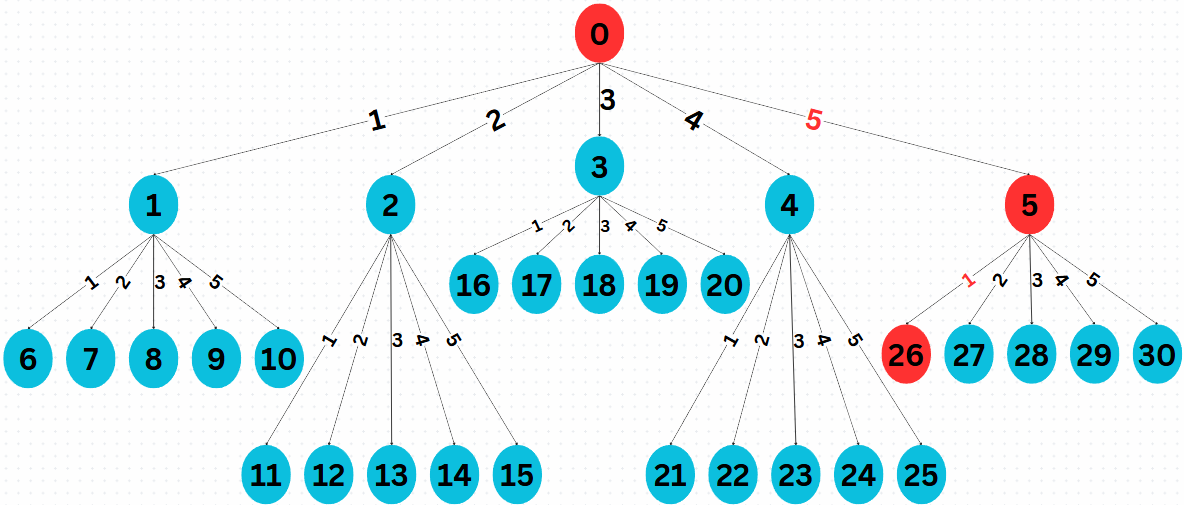}
    \caption{State tree in reverse matrix representation for searching the circuit in Fig. \ref{fig:bell_state_0}.}
    \label{fig:reversecase}
    \vspace{-0.2in}
\end{figure}

\subsubsection{Reverse Matrix Representation}

\begin{itemize}[leftmargin=*]
\item Actions
$\mathcal{A}^{-1} = \{H_0^{-1},H_1^{-1},T_0^{-1},T_1^{-1},\text{CNOT}_{01}^{-1}\}$, since $H^{-1}$ and $T^{-1}$ can be executed on either $q_0$ or $q_1$. An action $a \in \mathcal{A}^{-1}$ is represented as a matrix \(\mathit{A^{-1}}\in \mathbb{C}^{4\times 4}\). 



\item State space $\mathcal{S}$: The initial state is $S_0^{-1} = U$ given in (\ref{eq:target_matrix}) and the terminal state is $I_4$. 
Let $S^{-1}$ be the current state (a node in Fig. \ref{fig:reversecase}), $A^{-1} \in \mathcal{A}^{-1}$ be the action, then the resulting state at a child node $S'^{-1}$ is given by
\begin{equation}
   S^{'-1} = A^{-1} \cdot S^{-1}. 
\end{equation}
The state space $\mathcal{S}^{-1}$ is a tree in Fig. \ref{fig:reversecase}. The connecting lines 1, 2, 3, 4, and 5 correspond to the five actions in $\mathcal{A}^{-1}$. At initial state $S_0 = U$, taking an action $a \in \mathcal{A}^{-1}$ will generate $5$ states $\{S_1^{-1}, S_2^{-1}, S_3^{-1}, S_4^{-1}, S_5^{-1}\}$. Then, taking a second action $a \in \mathcal{A}^{-1}$ at a state $S^{-1} \in \{S_1^{-1}, S_2^{-1}, S_3^{-1}, S_4^{-1}, S_5^{-1}\}$ will generate 25 states $\{S_6^{-1}, S_7^{-1}, \ldots, S_{30}^{-1}\}$. Thus, $\mathcal{S}^{-1}$ has a total of 31 states.


\item Reward function $R$: At state $S_5^{-1}$, taking action $a = H^{-1}_0$, the reward $R(s = S_5^{-1}, a = H^{-1}_0)= 100$; otherwise, $R(s, a)=0$. 
\end{itemize}

\noindent\textbf{Example for Fig. \ref{fig:bell_state_0}}: Given initial state $S_0^{-1} = U$ in (\ref{eq:target_matrix}), we consider the optimal trajectory $S_0^{-1} \rightarrow S_5^{-1} \rightarrow S_{26}^{-1}$. \\
\noindent\textbf{State after taking the first action} $a = \text{CNOT}_{01}^{-1}$,
\begin{equation}
\begin{split}
& S_5^{-1} = \text{CNOT}_{01}^{-1} \cdot S_0^{-1} \\
& =\begin{pmatrix}
  1 & 0 & 0 & 0 \\
  0 & 1 & 0 & 0 \\
  0 & 0 & 0 & 1 \\
  0 & 0 & 1 & 0 
  \end{pmatrix}
  \frac{1}{\sqrt{2}}
  \begin{pmatrix}
  1 & 0 & 1 & 0 \\
  0 & 1 & 0 & 1 \\
  0 & 1 & 0 & -1 \\
  1 & 0 & -1 & 0 
  \end{pmatrix} 
  \\ & 
  = \frac{1}{\sqrt{2}}
  \begin{pmatrix}
  1 & 0 & 1 & 0 \\
  0 & 1 & 0 & 1 \\
  1 & 0 & -1 & 0 \\
  0 & 1 & 0 & -1 
  \end{pmatrix}.
\end{split}
\end{equation}

\noindent\textbf{Final state after taking the second action} $a = H_{0}^{-1}$,
\begin{equation}
\begin{split}
& S^{-1}_{26} = (H_0^{-1}\otimes I) S_5^{-1} \\
& =\frac{1}{2}\begin{pmatrix}
  1 & 0 & 1 & 0 \\
  0 & 1 & 0 & 1 \\
  1 & 0 & -1 & 0 \\
  0 & 1 & 0 & -1 
  \end{pmatrix}
  \begin{pmatrix}
  1 & 0 & 1 & 0 \\
  0 & 1 & 0 & 1 \\
  1 & 0 & -1 & 0 \\
  0 & 1 & 0 & -1 
  \end{pmatrix} = 
  I_4.
\end{split}
\end{equation}

To construct the target circuit, one can reverse the ordering of actions and take the inverse of each action. In this example, gate CNOT$_{01}^{-1}$ is followed by gate $H_{0}^{-1}$. Therefore, the result is $H_{0}$ followed by CNOT$_{01}$, which corresponds to the target circuit in Fig. \ref{fig:bell_state_0}.

\begin{table}[t]
\centering
\begin{tabular}{c|c|c|c|c|c}
\toprule[1pt]
States & $H_0$ & $H_1$ & $T_0$ & $T_1$ & CNOT$_{01}$ \\
\hline
0 & 90 & 0 & 0 & 0 & 0 \\
1 & 0 & 0 & 0 & 0 & 100 \\
2 & 0 & 0 & 0 & 0 & 0 \\
\vdots & 0 & 0 & 0 & 0 & 0 \\
30 & 0 & 0 & 0 & 0 & 0 \\
\bottomrule[0.8pt]
\end{tabular}
\caption{The learned Q-table for Bell state $|\Phi^+\rangle$.}
\label{tab:q_table_values}
\vspace{-0.1in}
\end{table}

\subsubsection{Tensor Network Representation}
\begin{figure}
    \centering
    \includegraphics[width=0.75\linewidth]{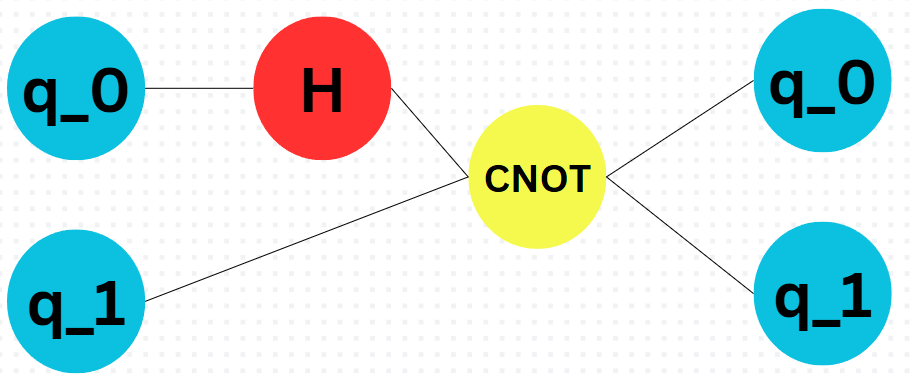}
    \caption{TN representation of Fig. \ref{fig:bell_state_0}.
    }
    \label{fig:TN_BELLE}
\end{figure}

 The Tensor Network (TN) is a powerful representation for quantum circuits. A tensor network is a collection of interconnected tensors. A single-qubit gate can be represented as a 2-order tensor, while a double-qubit gate can be represented as a 4-order tensor. For example, we convert the circuit in Fig. \ref{fig:bell_state_0} to Fig. \ref{fig:TN_BELLE}.

\begin{table}[t]
\centering
\begin{tabular}{c|c|c|c|c|c}
\toprule[0.9pt]
States & $H^{-1}_0$ & $H^{-1}_1$ & $T^{-1}_0$ & $T^{-1}_1$ & CNOT$_{01}^{-1}$ \\ 
\hline 
0 & 0 & 0 & 0 & 0 & 90 \\
\vdots & 0 & 0 & 0 & 0 & 0 \\
5 & 100 & 0 & 0 & 0 & 0 \\
\vdots & 0 & 0 & 0 & 0 & 0 \\
30 & 0 & 0 & 0 & 0 & 0 \\
\bottomrule[0.8pt]
\end{tabular}
\caption{The learned Q-table of reverse representation for Bell state $|\Phi^+\rangle$.}
\label{tab:q_table_values_r}
\vspace{-0.1in}
\end{table}

\begin{figure}
    \centering
    \includegraphics[width=1\linewidth]{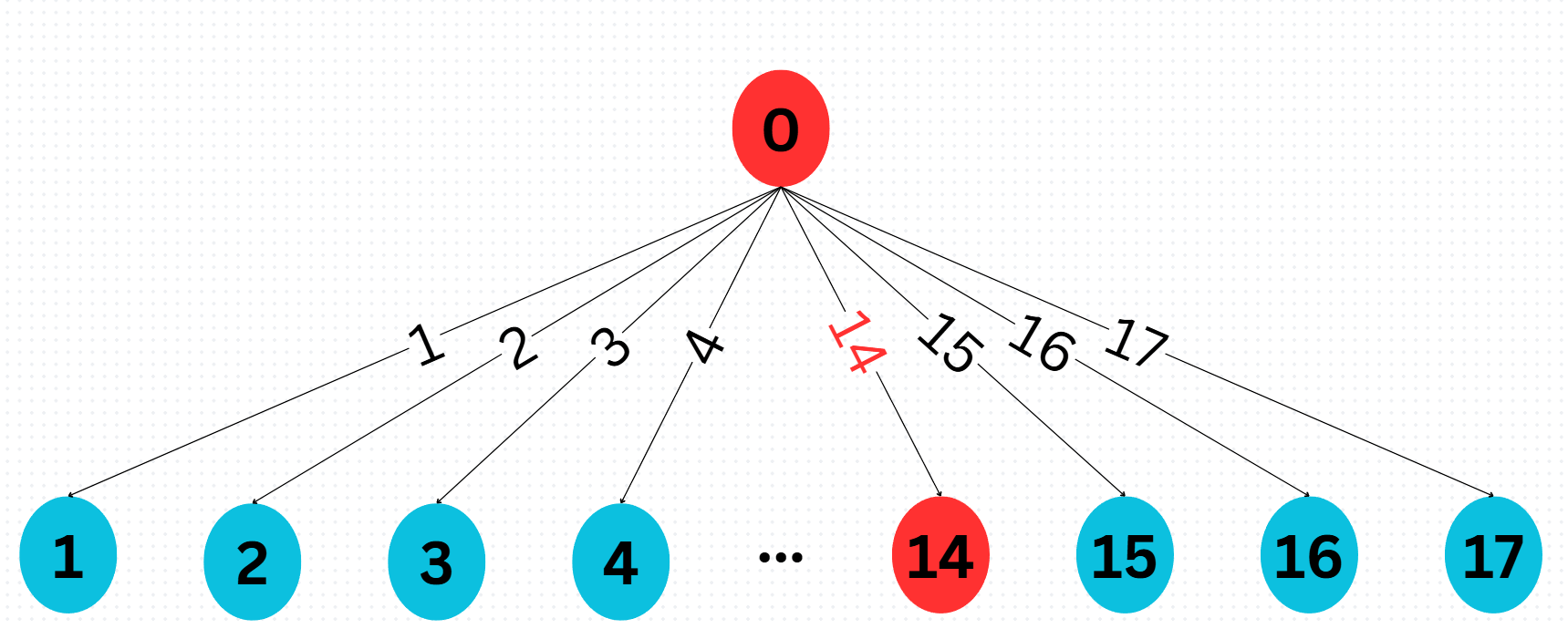}
    \caption{State tree in TN representation for searching the circuit in Fig. \ref{fig:bell_state_0}.}
    \label{fig:TN_BELL}
    \vspace{-0.2in}
\end{figure}




Consider Fig. \ref{fig:bell_state_0} and a universal gate set $G = \{H, T, \text{CNOT}_{01}\}$. The gate list is $L = \{H_0,H_1, T_0,T_1, \text{CNOT}\}$. We allow up to two gates for demonstration purpose. 
\begin{itemize}[leftmargin=*]
\item Actions $\mathcal{A}$ = $\{H_0, H_1, T_0, T_1, \text{CNOT}_{01}, \\
(H_0, H_1), (H_0, T_1), (H_1, T_0), (T_0, T_1),\\
(T_0, \text{CNOT}_{01}), (\text{CNOT}_{01}, T_0), (T_1, \text{CNOT}_{01}), (\text{CNOT}_{01}, T_1),\\
 (H_0, \text{CNOT}_{01}), (\text{CNOT}_{01}, H_0), (H_1, \text{CNOT}_{01}), (\text{CNOT}_{01}, H_1)\}$.
There are 17 different actions in total.
Taking action \( (H_0, \text{CNOT}_{01}) \) results in the TN representation in Fig. \ref{fig:TN_BELLE}.

\item State space $\mathcal{S}$: The initial state is \( S_0 = \ket{00} \), and the terminal state is $\left|\Phi^+\right\rangle$ given in (\ref{eq:target_state111}). Let \( S \) be the current state (a node in Fig.~\ref{fig:TN_BELL}), \( A \in \mathcal{A} \) be an action, then the resulting state at a child node \( S^{'} \) is given by:
\begin{equation}
   S^{'} = A \cdot S. 
\end{equation}

The state space \( \mathcal{S} \) is represented as a tree in Fig.~\ref{fig:TN_BELL}. The connecting lines \( 1, 2, 3, \dots, 17 \) correspond to the 17 actions in \( \mathcal{A} \). At the initial state \( S_0 = \ket{00} \), taking an action \( A \in \mathcal{A} \) will generate \( 17 \) states $\{S_1, S_2, S_3, \dots, S_{17}\}$. 
Thus, \( \mathcal{S} \) contains a total of \( 18 \) states.

\item Reward function $R$: At state $S_0$, taking action $a = (H_0, \text{CNOT}_{01})$, the reward $R(s = S_0, a = (H_0,\text{CNOT}_{01})) = 100$; otherwise, $R(s, a)=0$.

\end{itemize}
\noindent\textbf{Example for Fig. \ref{fig:bell_state_0}}: Given the initial state \( S_0 = \ket{00} \), we consider the optimal trajectory \( S_0 \rightarrow S_{14} \).



\noindent\textbf{State after the action:} \(a=(H_0, \text{CNOT}_{01})\)

\begin{equation}
\begin{split}
S_{14} &= \text{CNOT}_{01} \cdot (H \otimes I) \cdot S_0 \\
&= \text{CNOT}_{01} \cdot \left( \frac{1}{\sqrt{2}} \left( \ket{00} + \ket{10} \right) \right) \\
&= \frac{1}{\sqrt{2}} \left( \ket{00} + \ket{11} \right).
\end{split}
\end{equation}
which corresponds to the target circuit in Equation (\ref{eq:target_matrix}).

\section{Q-Learning and DQN Algorithms}

\subsection{Q-Learning Algorithm}

The Q-learning algorithm updates a Q-table \cite{watkins1992q} in each step as follows
\begin{equation}
\label{eq:Q}
\scalebox{0.66}{
$
\begin{split}
&Q^{\text{new}}(S_t, A_t) \leftarrow \underbrace{(1 - \alpha)}_{\textstyle\text{learning rate}} \cdot \underbrace{Q(S_t, A_t)}_{\textstyle\text{current value}} \\
& + \underbrace{\alpha}_{\textstyle\text{learning rate}} \cdot \left( \underbrace{R_{t+1}}_{\textstyle\text{reward}} + \underbrace{\gamma}_{\textstyle\text{discount factor}} \cdot \underbrace{\max_{a} Q(S_{t+1}, a)}_{\textstyle\text{estimate of optimal future value}} \right).
\end{split}$
}
\end{equation}

\subsubsection{Q-Table for Bell State $\left|\Phi^+\right\rangle$}

\begin{table}[t]
\centering
\begin{tabular}{c|c|c|c|c|c|c}
\toprule[1pt]
States & $A_1$ & $A_2 $& \dots & $A_{14}$ & \dots & $A_{17}$ \\
\hline
0 & 0 & 0 & \dots & 100 & \dots & 0 \\
1 & 0 & 0 & \dots & 0 & \dots & 0 \\
2 & 0 & 0 & \dots & 0 & \dots & 0 \\
\vdots & 0 & 0 & \dots & 0 & \dots & 0 \\
17 & 0 & 0 & \dots & 0 & \dots & 0 \\
\bottomrule[0.8pt]
\end{tabular}
\caption{The learned Q-table of TN representation for Bell state $\ket{\Phi}^+$.}
\label{tab:q_table_valuesTN}
\begin{flushleft}
\end{flushleft}
\vspace{-0.2in}
\end{table}
The rows of the Q-table in Table 1 and Table 2 correspond to 31 states in Fig. \ref{fig:bell_state_1} and Fig. \ref{fig:reversecase}, and the columns for five actions in $\mathcal{A}$ and $\mathcal{A}^{-1}$  for Matrix and Reverse Matrix Representation.
The rows of the Q-table in Table 3 correspond to 18 states in Fig. \ref{fig:TN_BELL}, and the columns for 17 actions in $\mathcal{A}$ for TN Representation.
The Q-table is initialized to all zeros and updated by (\ref{eq:Q}).
After 500 iterations, the results are given in Table \ref{tab:q_table_values}, Table \ref{tab:q_table_values_r}, and Table \ref{tab:q_table_valuesTN}, respectively. Each entry represents the expected return for taking an action in a given state. The parameters are as follows: learning rate $\alpha = 0.5$, reward for reaching the target circuit ${R} = 100$, discount factor $\gamma = 0.9$, and exploration rate $\epsilon =0.2$.

Using Table \ref{tab:q_table_values}, at initial state $S_0$, we take action $a = H_0$ and obtain state $S_1$. At state $S_1$, we take  action $a = \text{CNOT}_{01}$ and reach the target circuit in Fig. \ref{fig:bell_state_0}.



Using Table \ref{tab:q_table_values_r}, at initial state $S^{-1}_0$, we take action $a = \text{CNOT}_{01}^{-1}$ and obtain state $S^{-1}_5$. At state $S^{-1}_5$, we take action $a = H_0^{-1}$ and reach the target state $I_4$. By reversing the action ordering and taking the inverse of each action, the target circuit in Fig. \ref{fig:bell_state_0} is obtained.

Using Table \ref{tab:q_table_valuesTN}, at initial state \( S_0  \), we take action $a = \{H_0, \text{CNOT}_{01}\}$ and obtain the target state \( S_{14}  \) in Fig. \ref{fig:bell_state_0}.



\subsection{DQN Algorithm}
Deep Q-Network (DQN) method \cite{mnih2013playing} uses a neural network to approximate the Q-values for each state-action pair. The DQN algorithm utilizes two neural networks:
\begin{itemize}[leftmargin=*]
\item{Policy network} with parameter $\theta$: It consists of three fully connected layers, each with 128 neurons. The input is the state and the outputs are Q-values for each action.
\item{Target network} with parameter $\overline{\theta}$: A separate network that stabilizes the training process. It is periodically updated using $\overline{\theta} = (1-\alpha) \overline{\theta} + \alpha \theta$, where $\alpha$ is the learning rate.
\end{itemize}
 
Experiences stored in the replay buffer are randomly sampled to train the policy network, reducing correlations between consecutive samples. The loss function is defined as the Mean Squared Error (MSE) between the predicted Q-values from the policy network and the target Q-values \cite{mnih2013playing}:
\begin{equation}
\mathcal{L}_{\theta} = \text{MSE} \left(Q(s, a \mid \theta), R + \gamma \cdot \max_{a'} Q(s', a' \mid \overline{\theta})\right),
\end{equation}
where $Q(s, a \mid \theta)$ denotes the Q-value predicted by the policy network for the current state-action pair, and the target Q-value is calculated as the immediate reward $R$ plus the discounted maximum next-step Q-value $\max_{a'} Q(s', a' \mid \overline{\theta})$, which is estimated using the target network. 
The parameters are as follows:  $\alpha = 0.1$, $\gamma = 0.95$,  batch size $= 64$, replay buffer size $= 10000$, and max gate count $= 20$.

\section{Experiment Results}
%

 \begin{figure}[h]
    \centering
    \scalebox{0.75}{ 
\begin{quantikz}
    q_0 :~& \targ{} & \ghost{P^\dagger} \\
    q_1 :~& \ctrl{-1} & \ghost{P^\dagger} \\
    q_2 :~& \ctrl{-2} & \ghost{P^\dagger}
\end{quantikz}
\quad = \quad
\begin{quantikz}
    & \gate{H} & \gate{P} & \qw         & \gate{P^\dagger} & \qw        & \gate{P} & \gate{H} & \ghost{P^\dagger} \\
    & \qw & \ctrl{-1}     & \targ{}     & \ctrl{-1}        & \targ{}    & \qw      & \qw      & \ghost{P^\dagger} \\
    & \qw      & \qw      & \ctrl{-1}   & \qw              & \ctrl{-1}  & \ctrl{-2} & \qw     & \ghost{P^\dagger}
\end{quantikz}

    }

    \caption{A design of Toffoli Gate with fewer gates.}
    \label{fig:toff1}
    \vspace{-0.1in}
\end{figure}
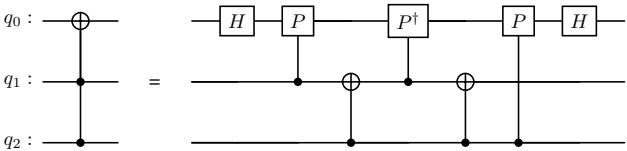


We verify the above three MDP modelings for $10$ well-known quantum circuits, namely, circuits to generate 4 Bell states, SWAP gate, iSWAP gate, CZ gate, GHZ Gate, Z gate, and Toffoli gate. For Matrix and Reverse Matrix Representations, we apply both Q-learning and DQN algorithms, while for TN Representation, we applied only Q-learning. Our codes can be found at this link\footnote{\url{https://github.com/YangletLiu/CSCI4961_labs_projects/tree/main}}.


\textbf{Toffoli Gate}: we used an action set with gates in Fig. \ref{fig:toff1},
\[
\mathcal{A} = \{\text{CNOT}_{21}, H_0, \text{CP}_{10}, \text{CP}^{-1}_{10}, \text{CP}_{20}\},
\]
where the $\text{CP}$ gate refers to controlled-phase gate with a phase shift of $\frac{\pi}{2}$, and $\text{CP}^{-1}$ with a phase shift of $-\frac{\pi}{2}$.

An expert trajectory were stored in the replay buffer to improve learning efficiency.
For Matrix Representation, the expert trajectory is:
\[  \{ H_0 \rightarrow \text{CP}_{10} \rightarrow \text{CNOT}_{21} \rightarrow \text{CP}^{-1}_{10} \rightarrow \]\[\text{CNOT}_{21} \rightarrow \text{CP}_{20} \rightarrow H_0 \}.
\]

\noindent For Reverse Matrix Representation,
\[
\mathcal{A}^{-1} = \{\text{CNOT}_{21}^{-1}, H_0^{-1}, \text{CP}_{10}^{-1}, \text{CP}_{10}, \text{CP}_{20}^{-1}\},
\]
and the expert trajectory becomes to:
\[  \{ H_0^{-1} \rightarrow \text{CP}_{20}^{-1} \rightarrow \text{CNOT}_{21}^{-1} \rightarrow \text{CP}_{10} \rightarrow\]\[ \text{CNOT}_{21}^{-1} \rightarrow \text{CP}_{10}^{-1} \rightarrow H_0^{-1} \}.
\]




Each state expands in a branching factor (size of actions) \( c \) across \( b + 1 \) levels (length of the tasks+1), as in Fig. \ref{fig:bell_state_1},  the size of the state space is given by a geometric series:
\begin{equation}
  \text{Size of state space} 
  = c^0 + c^1 + \cdots + c^b 
  = \frac{c^{b+1} - 1}{c - 1}.
\end{equation}
The complexity of the task is measured by the size of the states space, given in Table~\ref{tab:TaskDifficulty}. To evaluate the effectiveness of Q-learning and DQN, we conduct 100 rounds. In each round, the agent is trained for 100 episodes, and we measure the success ratio (in percentage) of correct testing results over the 100 rounds. The results are summarized in Table~\ref{tab:P_result}.


From Table~\ref{tab:P_result}, we observe that both Q-learning and DQN perform well on simpler tasks, such as generating the Bell state \( |\Phi^+\rangle \). However, as task complexity increases, for example, the iSWAP gate task with a state space size of \( 5^6 \), the performance of both algorithms significantly degrades, indicating the challenges of learning in large state spaces.

\section{Conclusion and Future Work}

In this paper, we applied Q-learning and Deep Q-Network (DQN) algorithms to three MDP modelings of the quantum circuit design task. We demonstrated that RL algorithms successfully discovered the expected quantum circuits for $4$ Bell states, SWAP gate, iSWAP gate, CZ gate, GHZ gate, Z gate, and Toffoli gate. We noticed that Reverse Matrix Representation and TN Representation have greater potential in this problem. For more difficult tasks, both Q-learning and DQN struggle to converge due to insufficient sampling quality and efficiency. 


In future work, we will improve sample quality and implement algorithms like Monte Carlo Tree Search (MCTS) to increase efficiency and address the convergence challenge. Finally, we will investigate the robustness of RL algorithms by testing more complex quantum circuits.

\newpage
\bibliography{ref}

\begin{thebibliography}{9}
\providecommand{\natexlab}[1]{#1}

\bibitem[{Ali et~al.(2015)Ali, Hirayama, Yamanaka, and Nishitani}]{ali2015quantum}
Ali, M.~B.; Hirayama, T.; Yamanaka, K.; and Nishitani, Y. 2015.
\newblock Quantum cost reduction of reversible circuits using new Toffoli decomposition techniques.
\newblock In \emph{International Conference on Computational Science and Computational Intelligence (CSCI)}, 59--64. IEEE.

\bibitem[{Arute et~al.(2019)Arute, Arya, Babbush, Bacon, Bardin, Barends, Biswas, Boixo, Brandao, Buell et~al.}]{arute2019quantum}
Arute, F.; Arya, K.; Babbush, R.; Bacon, D.; Bardin, J.~C.; Barends, R.; Biswas, R.; Boixo, S.; Brandao, F.~G.; Buell, D.~A.; et~al. 2019.
\newblock Quantum supremacy using a programmable superconducting processor.
\newblock \emph{Nature}, 574(7779): 505--510.

\bibitem[{Bhat, Khanday, and Shah(2022)}]{bhat2022optimal}
Bhat, H.~A.; Khanday, F.~A.; and Shah, K.~A. 2022.
\newblock Optimal quantum circuit decomposition of reversible gates on IBM quantum computer.
\newblock In \emph{International Conference on Multimedia, Signal Processing and Communication Technologies (IMPACT)}, 1--4. IEEE.

\bibitem[{Gill et~al.(2021)Gill, Kumar, Singh, Singh, Kaur, Usman, and Buyya}]{gill2021quantumcomputingtaxonomysystematic}
Gill, S.~S.; Kumar, A.; Singh, H.; Singh, M.; Kaur, K.; Usman, M.; and Buyya, R. 2021.
\newblock Quantum Computing: A Taxonomy, Systematic Review and Future Directions.
\newblock arXiv:2010.15559.

\bibitem[{Liu and Zhang(2023)}]{NEURIPS2023_d41b7001}
Liu, X.-Y.; and Zhang, Z. 2023.
\newblock Classical Simulation of Quantum Circuits: Parallel Environments and Benchmark.
\newblock In \emph{Advances in Neural Information Processing Systems}, volume~36, 67082--67102.

\bibitem[{Meirom et~al.(2022)Meirom, Maron, Mannor, and Chechik}]{meirom2022optimizing}
Meirom, E.; Maron, H.; Mannor, S.; and Chechik, G. 2022.
\newblock Optimizing tensor network contraction using reinforcement learning.
\newblock In \emph{International Conference on Machine Learning}, 15278--15292. PMLR.

\bibitem[{Mnih et~al.(2013)Mnih, Kavukcuoglu, Silver, Graves, Antonoglou, Wierstra, and Riedmiller}]{mnih2013playing}
Mnih, V.; Kavukcuoglu, K.; Silver, D.; Graves, A.; Antonoglou, I.; Wierstra, D.; and Riedmiller, M. 2013.
\newblock Playing Atari with deep reinforcement learning.
\newblock \emph{arXiv preprint arXiv:1312.5602}.

\bibitem[{Sogabe et~al.(2022)Sogabe, Kimura, Chen, Shiba, Kasahara, Sogabe, and Sakamoto}]{sogabe2022model}
Sogabe, T.; Kimura, T.; Chen, C.-C.; Shiba, K.; Kasahara, N.; Sogabe, M.; and Sakamoto, K. 2022.
\newblock Model-free deep recurrent Q-network reinforcement learning for quantum circuit architectures design.
\newblock \emph{Quantum Reports}, 4(4): 380--389.

\bibitem[{Watkins and Dayan(1992)}]{watkins1992q}
Watkins, C.~J.; and Dayan, P. 1992.
\newblock Q-learning.
\newblock \emph{Machine Learning}, 8(3-4): 279--292.

\end{thebibliography}

\newpage

\section*{Appendix: Task Description}
\begin{table*}[ht!]
\centering
\begin{tabular}{@{}lcccccl@{}}
\toprule
\textbf{Task Name} & \textbf{Qubits} & \textbf{Actions} & \textbf{Length} & \textbf{Space Size} & \textbf{Universal
 Gate Set} \\ \midrule

Bell state $|\Phi^+\rangle$ & 2 & 6 & 2 & $43$ & $\{H, \text{CNOT}, T\}$ \\
Bell state $|\Phi^-\rangle$ & 2 & 6 & 3 & $259$ & $\{H, \text{CNOT}, T, X\}$ \\
Bell state $|\Psi^+\rangle$ & 2 & 6 & 3 & $259$ & $\{H, \text{CNOT}, T, X\}$ \\
Bell state $|\Psi^-\rangle$ & 2 & 8 & 5 & $37449$ & $\{H, \text{CNOT}, T, X, Z\}$ \\
SWAP gate & 2 & 6 & 3 & $259$ & $\{H, \text{CNOT}, T\}$ \\
iSWAP gate & 2 & 6 & 5 & $9331$  & $\{H, \text{CNOT}, T\}$ \\
CZ gate & 2 & 6 & 3 & $259$ & $\{H, \text{CNOT}, T\}$ \\
GHZ gate & 3 & 8 & 3 & $585$ &$\{H, \text{CNOT}, T\}$ \\
Z gate & 3 & 10 & 2 & $111$ & $\{H, \text{CNOT}, T, S\}$ \\
Toffoli gate & 3 & 5 & 7 & $97656$ &  Special Case \\

\bottomrule
\end{tabular}
\caption{Task descriptions. \\
}
\label{tab:TaskDifficulty}
\begin{flushleft}
\end{flushleft}

\end{table*}

\begin{table}[h!]
\begin{tabularx}{\columnwidth}{@{}lcr@{}}
\toprule
\textbf{Task Name}         & \textbf{Detailed Action Set} \\
\midrule
Bell state $|\Phi^+\rangle$ & $\{H_0, H_1, T_0, T_1, \text{CNOT}_{01}, \text{CNOT}_{10}\}$                                                                \\

Bell state $|\Phi^-\rangle$ & $\{H_0, H_1, T_0, X_0, X_1, \text{CNOT}_{01}\}$                                                       \\

Bell state $|\Psi^+\rangle$ & $\{H_0, H_1, T_0, X_0, X_1, \text{CNOT}_{01}\}$                                                       \\

Bell state $|\Psi^-\rangle$ & $\{H_0, H_1, T_0,  X_0, X_1, Z_0, Z_1, \text{CNOT}_{01}\}$                                            \\

SWAP gate                   & $\{H_0, H_1, T_0, T_1, \text{CNOT}_{01}, \text{CNOT}_{10}\}$                                                                 \\

iSWAP gate                  & $\{H_0, H_1, T_0, T_1, \text{CNOT}_{01}, \text{CNOT}_{10}\}$                                                                \\

CZ gate                     & $\{H_0, H_1, T_0, T_1, \text{CNOT}_{01}, \text{CNOT}_{10}\}$                                                                 \\

GHZ gate                    & $\{H_0, H_1, H_2, T_0, T_1, T_2, \text{CNOT}_{01}, \text{CNOT}_{12} \}$  \\

Toffoli gate                & $\{\text{CNOT}_{21}, H_0, \text{CP}_{10}, \text{CP}^{-1}_{10}, \text{CP}_{20}\}$ \\                             
\bottomrule
\end{tabularx}
\caption{Detailed action sets.}
\label{tab:ExpandedActionSet}
\end{table}

\subsection*{Two-Qubits Action Set (TN)}
\noindent For tasks below in TN representation:
\begin{itemize}
    \item Bell state \( |\Phi^-\rangle \),
    \item Bell state \( |\Psi^+\rangle \),
    \item Bell state \( |\Psi^-\rangle \),

\end{itemize}
They share the same action set:

\noindent$\{
H_0, H_1, T_0, T_1, X_0, X_1, \text{CNOT}_{01}, \\
(H_0, H_1), (H_0, T_1), (H_1, T_0), (T_0, T_1), (Z_0, Z_1), \\
(T_0, \text{CNOT}_{01}), (\text{CNOT}_{01}, T_0), (T_1, \text{CNOT}_{01}), (\text{CNOT}_{01}, T_1), \\
(H_0, \text{CNOT}_{01}), (\text{CNOT}_{01}, H_0), (H_1, \text{CNOT}_{01}), (\text{CNOT}_{01}, H_1)
\}$\\

\noindent For tasks below in TN representation:
\begin{itemize}
    \item SWAP gate,
    \item iSWAP gate,
    \item CZ gate,

\end{itemize}
They share the same action set:

\noindent$\{
H_0, H_1, T_0, T_1, \text{CNOT}_{01}, \text{CNOT}_{10},(H_0, H_1), (H_0, T_1), \\
 (H_1, T_0), (T_0, T_1), (\text{CNOT}_{01}, \text{CNOT}_{10}), (\text{CNOT}_{10}, \text{CNOT}_{01}), \\
(T_0, \text{CNOT}_{01}), (\text{CNOT}_{01}, T_0), (T_1, \text{CNOT}_{01}), (\text{CNOT}_{01}, T_1), \\
(H_0, \text{CNOT}_{01}), (\text{CNOT}_{01}, H_0), (H_1, \text{CNOT}_{01}), (\text{CNOT}_{01}, H_1)
\}$\\

\subsection*{Three-Qubit Action Set (TN)}
For tasks below in TN representation:
\begin{itemize}
    \item GHZ gate,
    \item Z gate,
\end{itemize}
They share the same action set:

\noindent$\{
H_0, H_1,H_2, T_0, S_0,S_1,S_2,T_1,T_2,\text{CNOT}_{01},\text{CNOT}_{12},\text{CNOT}_{02},\\
(H_0, H_1), (H_0, T_1), (T_0, H_1), (T_0, T_1),\\
(H_0, H_2), (H_0, T_2), (T_0, H_2), (T_0, T_2),\\
(H_1, H_2), (H_1, T_2), (T_1, H_2), (T_1, T_2),\\
(H_0, \text{CNOT}_{01}), (T_0, \text{CNOT}_{01}), (H_1, \text{CNOT}_{01}), \\
(T_1, \text{CNOT}_{01}), (H_2, \text{CNOT}_{01}), (T_2, \text{CNOT}_{01}),\\
(H_0, \text{CNOT}_{02}), (T_0, \text{CNOT}_{02}), (H_1, \text{CNOT}_{02}), \\
(T_1, \text{CNOT}_{02}), (H_2, \text{CNOT}_{02}), (T_2, \text{CNOT}_{02}),\\
(H_0, \text{CNOT}_{12}), (T_0, \text{CNOT}_{12}), (H_1, \text{CNOT}_{12}), \\
(T_1, \text{CNOT}_{12}), (H_2, \text{CNOT}_{12}), (T_2, \text{CNOT}_{12}),\\
(\text{CNOT}_{01}, \text{CNOT}_{02}), (\text{CNOT}_{01}, \text{CNOT}_{12}), 
(\text{CNOT}_{02}, \text{CNOT}_{12})
\}$
\section*{Appendix: Test Result}
\begin{table*}[]
    \centering
\begin{tabular}{cccccc}
      \toprule[1pt]
      \textbf{Gates} & \textbf{Q-Learning} &\textbf{Q-Learning (Reverse)}& \textbf{DQN}&\textbf{DQN (Reverse)}&\textbf{Q-Learning (TN)} \\
    \hline
      Bell state $|\Phi^+\rangle$ & $86\%$ & $85\%$&$33\%$& $39\%$ & \textbf{100\%}\\
      Bell state $|\Phi^-\rangle$ & $41\%$ & $25\%$ & $18\%$ & $20\%$ & \textbf{94\%}\\
      Bell state $|\Psi^+\rangle$ & $55\%$ & $53\%$ & $21\%$ & $17\%$ &  \textbf{95\%}\\
      Bell state $|\Psi^-\rangle$ & $5\%$ & $4\%$ & $6\%$ & $4\%$& \textbf{15\%}\\
      SWAP gate & $10\%$ &$15\%$&$21\%$& \textbf{27\%} & $3\%$ \\
      iSWAP gate & $2\%$ & $1\%$ &$2\%$&\textbf{5\%} & $2\%$ \\
      CZ gate & $69\%$ & \textbf{77\%} &$16\%$& $17\%$ & $19\%$ \\
      GHZ gate & $34\%$ & $17\%$ & $13\%$ & $20\%$ &\textbf{45\%} \\
      Z gate & \textbf{50\%} & $38\%$ & $17\%$ & $19\%$ &$13\%$ \\
          Toffoli gate& $87\%$ & \textbf{91\%} & $1\%$ & $3\%$ &- \\
      \bottomrule[1pt]
    \end{tabular}
    \caption{Success ratios (in percentage) over $100$ training rounds, respectively.}
    \label{tab:P_result}
\end{table*}

\section*{Appendix: Reward Calculation}

The reward for all tasks in Q-Learning, Q-Learning (Reverse), 
 DQN, DQN (Reverse); and four tasks in TN Representation (SWAP gate, iSWAP gate, CZ gate, Z gate) are calculated as follows:

\begin{enumerate}
    \item The quantum circuit is executed using a Qiskit simulator.
    \item The unitary operator of the current circuit is compared with the target unitary operator.
    \item If the comparison results in a value greater than 0.99, a reward of 100 is given.
\end{enumerate}

The reward calculation can be expressed as:

\[
\text{Reward} = 
\begin{cases} 
100, & \text{if } \frac{\left| \text{Tr}(S'^\dagger U) \right|}{2^{\text{num\_qubits}}} > 0.99 \\
0, & \text{otherwise}
\end{cases}
\]

\noindent where \( S' \) is the current unitary operator and \( U \) is the target unitary operator.\\

The reward for the five tasks in TN representation (four Bell states, GHZ gate) are calculated as follows:

\begin{enumerate}
    \item The quantum state of the circuit is obtained using the function \texttt{\_get\_quantum\_state}.
    \item The current state is compared with the target state.
    \item If the comparison results in a value greater than 0.99, a reward of 100 is given.
\end{enumerate}

The reward calculation can be expressed as:

\[
\text{Reward} = 
\begin{cases} 
100, & \text{if } |\langle S' | U \rangle |^2 > 0.99 \\
0, & \text{otherwise}
\end{cases}
\]
where \( S' \) is the current state vector and \( U \) is the target state vector, and\( \langle S' | U \rangle \) represents the inner product between the current state and the target state.

\section*{Appendix: Examples of Quantum Circuits}

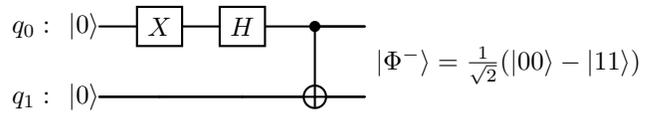
\begin{figure}[H]
\begin{tikzpicture}
    \node (circuit) [inner sep=0pt] {
        \begin{quantikz}
            q_0:~ \ket{0} & \gate{X} & \gate{H} & \ctrl{1} & \qw \\
            q_1:~ \ket{0} & \qw      & \qw      & \targ   & \qw & \qw
        \end{quantikz}
        
    };
    \node[right=0cm of circuit,yshift=-0.1cm] {$\ket{\Phi^-} = \frac{1}{\sqrt{2}}(\ket{00} - \ket{11})$};
\end{tikzpicture}
\caption{A quantum circuit to generate Bell state $\ket{\Phi^-}$.}
\label{fig:bell_state_2}
\end{figure}

\begin{figure}[H]
\centering
\begin{tikzpicture}
    \node (circuit) [inner sep=0pt] {
        \begin{quantikz}
            q_0:~ \ket{0}  & \gate{H} & \ctrl{1} & \qw \\
            q_1:~ \ket{0} & \gate{X}         & \targ   & \qw & \qw
        \end{quantikz}   
    };    
    \node[right=0cm of circuit,yshift=-0.1cm] {$\ket{\Psi^+} = \frac{1}{\sqrt{2}}(\ket{01} + \ket{10})$};
\end{tikzpicture}
\caption{A quantum circuit to generate Bell state $\ket{\Psi^+}$.}
\label{fig:bell_state_3}
\end{figure}

\begin{figure}[H]
\centering
\begin{tikzpicture}
    \node (circuit) [inner sep=0pt] {
        \begin{quantikz}
            q_0:~ \ket{0} & \gate{H} & \gate{Z} & \ctrl{1} & \qw \\
            q_1:~ \ket{0} & \gate{X} & \gate{Z}        & \targ   & \qw & \qw
        \end{quantikz}
        
    };
    \node[right=0cm of circuit,yshift=-0.1cm] {$\ket{\Psi^-} = \frac{1}{\sqrt{2}}(\ket{01} - \ket{10})$};
\end{tikzpicture}
\caption{A quantum circuit to generate Bell state $\ket{\Psi^-}$.}
\label{fig:bell_state_4}
\end{figure}

\begin{figure}[H]
\centering
\begin{tikzpicture}
    \node (circuit) [inner sep=0pt] {
        \begin{quantikz}
            q_0:~ \ket{0} & \targ{} & \ctrl{1}& \targ{}  & \qw \\
            q_1:~ \ket{0} & \ctrl{-1} & \targ{}& \ctrl{-1}  & \qw
        \end{quantikz}
    };
    \node[right=0cm of circuit, yshift=-0.1cm] {$\text{SWAP} =
    \begin{bmatrix}
    1 & 0 & 0 & 0 \\
    0 & 0 & 1 & 0 \\
    0 & 1 & 0 & 0 \\
    0 & 0 & 0 & 1
    \end{bmatrix}$};

\end{tikzpicture}
\caption{A quantum circuit to implement the SWAP gate and its matrix form.}
\label{fig:swap_gate}
\end{figure}

\begin{figure}[H]
\centering
\scalebox{0.92}{
\begin{tikzpicture}
    \node (circuit) [inner sep=0pt] {
        \begin{quantikz}
            q_0:~ \ket{0} & \ctrl{1} & \qw      & \qw       & \targ{} & \ctrl{1} & \qw \\
            q_1:~ \ket{0} & \targ{} & \gate{T}  & \gate{T}  & \ctrl{-1} & \targ{} & \qw
        \end{quantikz}
    };
    \node[right=0cm of circuit, yshift=-0.1cm] {$\text{iSWAP} =
    \begin{bmatrix}
    1 & 0 & 0 & 0 \\
    0 & 0 & i & 0 \\
    0 & i & 0 & 0 \\
    0 & 0 & 0 & 1
    \end{bmatrix}$};
\end{tikzpicture}
}
\caption{A quantum circuit to implement the iSWAP gate and its matrix form.}
\label{fig:iswap_gate}
\end{figure}

\begin{figure}[H]
\centering
\scalebox{0.92}{
\begin{tikzpicture}
    \node (circuit) [inner sep=0pt] {
        \begin{quantikz}
            q_0:~ \ket{0} & \gate{S}& \gate{H}& \ctrl{1}       & \targ{}& \qw  & \qw \\
            q_1:~ \ket{0} & \gate{S}& \qw & \targ{}   & \ctrl{-1} & \gate{H} & \qw
        \end{quantikz}
    };
        \node[right=0cm of circuit, yshift=-0.1cm] {$\text{iSWAP} =
    \begin{bmatrix}
    1 & 0 & 0 & 0 \\
    0 & 0 & i & 0 \\
    0 & i & 0 & 0 \\
    0 & 0 & 0 & 1
    \end{bmatrix}$};
\end{tikzpicture}
}
\caption{A quantum circuit to implement the iSWAP gate by Qiskit and its matrix form.\footnotemark}
\label{fig:iswap_gate_Qiskit}
\end{figure}
\footnotetext{\url{https://docs.quantum.ibm.com/api/qiskit/qiskit.circuit.library.iSwapGate}}

Our design (Fig. \ref{fig:iswap_gate}) differs from Qiskit's implementation (Fig. \ref{fig:iswap_gate_Qiskit}) in the following ways:
\begin{itemize}
    \item Our design uses $3$ CNOT gates, while Qiskit's design uses only $2$.
    \item Our design uses $4$ gates, while Qiskit's design used $6$, if we considered two $T$ gates are equivalent to one $S$ gate.
\end{itemize}

\begin{figure}[H]
\centering
\begin{tikzpicture}
    \node (circuit) [inner sep=0pt] {
        \begin{quantikz}
            q_0:~ \ket{0} & \gate{H} & \targ{} & \gate{H} & \qw \\
            q_1:~ \ket{0} & \qw      & \ctrl{-1}  & \qw      & \qw
        \end{quantikz}
    };
    \node[right=0cm of circuit, yshift=-0.1cm] {$\text{CZ} =
    \begin{bmatrix}
    1 & 0 & 0 & 0 \\
    0 & 1 & 0 & 0 \\
    0 & 0 & 1 & 0 \\
    0 & 0 & 0 & -1
    \end{bmatrix}$};
\end{tikzpicture}
\caption{A quantum circuit to implement the CZ gate and its matrix form.}
\label{fig:cz_gate}
\end{figure}
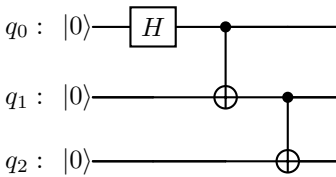

\begin{figure}[H]
\centering
\begin{tikzpicture}
    \node (circuit) [inner sep=0pt] {
        \begin{quantikz}
            q_0:~ \ket{0} & \gate{H} & \ctrl{1} & \qw      & \qw \\
            q_1:~ \ket{0} & \qw      & \targ{}  & \ctrl{1} & \qw \\
            q_2:~ \ket{0} & \qw      & \qw      & \targ{}  & \qw
        \end{quantikz}
    };
    \node[right=0cm of circuit, yshift=-0.1cm] {$\text{GHZ state} = \frac{1}{\sqrt{2}} (\ket{000} + \ket{111})$};
\end{tikzpicture}
\caption{A quantum circuit to generate a GHZ state.}
\label{fig:ghz_gate}
\end{figure}

\begin{figure}[H]
\centering
\begin{tikzpicture}
    \node (circuit) [inner sep=0pt] {
        \begin{quantikz}
            q_0:~ \ket{0} & \gate{S} & \gate{S} & \qw       \\
            q_1:~ \ket{0} & \qw      & \qw      & \qw       \\
            q_2:~ \ket{0} & \qw      & \qw      & \qw       
        \end{quantikz}
    };
    \node[right=0cm of circuit, yshift=-0.1cm] {$\text{Z} =
    \begin{bmatrix}
    1 & 0 & 0 & 0 & 0 & 0 & 0 & 0 \\
    0 & 1 & 0 & 0 & 0 & 0 & 0 & 0 \\
    0 & 0 & 1 & 0 & 0 & 0 & 0 & 0 \\
    0 & 0 & 0 & 1 & 0 & 0 & 0 & 0 \\
    0 & 0 & 0 & 0 & -1 & 0 & 0 & 0 \\
    0 & 0 & 0 & 0 & 0 & -1 & 0 & 0 \\
    0 & 0 & 0 & 0 & 0 & 0 & -1 & 0 \\
    0 & 0 & 0 & 0 & 0 & 0 & 0 & -1
    \end{bmatrix}$};
\end{tikzpicture}
\caption{A quantum circuit to implement the Z gate in 3-qubits space and its matrix form.}
\label{fig:z_gate_3qubits}
\end{figure}

\section*{Appendix: Q-learning and DQN Environment (Example for Two Qubits)}

Environments follow the training loop according to the example code snippet (Listing~\ref{training_loop}).

\subsection*{Q-Learning Environment}
\begin{itemize}
    \item \textbf{State Space}: 
    The environment consists of \textbf{100 discrete states}, each representing a unique configuration of the system.
    \item \textbf{Q-Table}: 
    A \textbf{$100 \times 6$ table} is used to store the Q-values for each state-action pair.
    \item \textbf{Training Parameters}:
    \begin{itemize}
        \item \textbf{Learning Rate} (\(\alpha\)): 0.1
        \item \textbf{Discount Factor} (\(\gamma\)): 0.95
        \item \textbf{Exploration Rate} (\(\epsilon\)): Initial value of 1.0, decays at a rate of 0.99,with a minimum value of 0.05.
    \end{itemize}
\end{itemize}

\subsection*{DQN Environment}
\begin{itemize}
    \item \textbf{State Space}: 
    The state is represented as a \textbf{feature vector} and passed to a neural network. The environment supports continuous state spaces.
    \item \textbf{Neural Network}: 
    The Q-values are approximated using a 3-layer fully connected neural network:
    \begin{itemize}
        \item Input Layer: Accepts the state vector as input.
        \item Two Hidden Layers: Each with \textbf{128 neurons} and ReLU activation.
        \item Output Layer: Produces Q-values for \textbf{6 actions}.
    \end{itemize}
    \item \textbf{Training Parameters}:
    \begin{itemize}
        \item \textbf{Learning Rate} (\(\alpha\)): 0.1
        \item \textbf{Discount Factor} (\(\gamma\)): 0.95
        \item \textbf{Exploration Rate} (\(\epsilon\)): Initial value of 0.9, decays at a rate of 0.995, with a minimum value of 0.05.
        \item \textbf{Batch Size}: 64
        \item \textbf{Replay Buffer Size}: 10,000
        \item \textbf{Target Network Update}: Every 100 episodes.
    \end{itemize}

\end{itemize}
\textbf{Note}: 
We designed our own custom environment \texttt{QuantumEnv} built with \texttt{gym}\footnote{For details on creating a custom \texttt{gym} environment, refer to the official documentation: \url{https://www.gymlibrary.dev/content/environment_creation/}}. The environment implementation can be found in this link. 
\footnote{\url{https://github.com/YangletLiu/CSCI4961_labs_projects/tree/main}}

\lstset{
  basicstyle=\ttfamily\footnotesize,
  keywordstyle=\color{blue}\bfseries,
  commentstyle=\color{gray}\itshape,
  stringstyle=\color{red},
  showstringspaces=false,
  numbers=left,
  numberstyle=\tiny,
  breaklines=true,
  frame=lines,
  rulecolor=\color{black},
  backgroundcolor=\color{lightgray!10},
  xleftmargin=15pt,
  framexleftmargin=10pt,
  language=Python,
  captionpos=b
}

\begin{table*}[ht]
\centering
\begin{lstlisting}[caption={Training loop for Q-learning and DQN.}, label={training_loop}]

def train_agent(agent, environment, episodes, max_steps_per_episode, method='Q'):
    for episode in range(episodes):
        # Reset the environment and initialize variables
        state_index = environment.reset()
        total_reward = 0

        for step in range(max_steps_per_episode):
            # Select an action based on the current state
            action, action_index = agent.choose_action(state_index)

            # Execute the action and observe the result
            next_state_index, reward, done = environment.step(action[0], action[1])
            total_reward += reward

            # Update logic based on the method
            if method == 'Q':  # Q-Learning update
                agent.update_q_table(state_index, action_index, reward, next_state_index)
            elif method == 'DQN':  # DQN logic
                agent.remember(state_index, action_index, reward, next_state_index, done)
                agent.replay()

            # Update the current state
            state_index = next_state_index

            # If the episode is done, exit the loop
            if done:
                break

        # Additional updates for DQN
        if method == 'DQN' and (episode + 1) % 100 == 0:
            agent.update_target_net()

\end{lstlisting}

\end{table*}

\section*{Appendix: Expert Trajectories for Toffoli Gate in Q-Learning and DQN}
\subsection*{Overview}
Both Q-Learning and DQN use an \textbf{expert action sequence} to embed optimal behavior for constructing a Toffoli gate. This sequence (Fig. \ref{fig:toff1}) guides the agent's learning process by providing predefined state-action pairs that achieve the desired result.

\subsection*{Similarities}
\begin{itemize}
    \item The expert trajectory is applied over multiple iterations, starting with an environment reset.
    \item Selected actions are executed sequentially, returning the next state, reward, and completion flag.
\end{itemize}

\subsection*{Differences}
\begin{itemize}
    \item \textbf{Q-Learning:} 
    \begin{itemize}
        \item Applied over \textbf{10 iterations} of the expert trajectory.
        \item The Q-Table is updated at the end of each iteration using the transitions observed during the trajectory.
    \end{itemize}
    \item \textbf{DQN:} 
    \begin{itemize}
        \item Applied over \textbf{150 iterations} of the expert trajectory.
        \item After every action in the trajectory, the transition is stored in memory, and a small replay step is performed.
        \item At the end of each iteration, the target network is updated using \texttt{update\_target\_net()} for stable training.
    \end{itemize}
\end{itemize}

\end{document}